\documentclass[aps,a4paper,twocolumn,floatfix,showkeys,
groupaddress]{revtex4}

\usepackage{bm,amsmath,amssymb,amsfonts}

\usepackage[dvips]{graphicx}
\usepackage[colorlinks=true,citecolor=magenta,urlcolor=red,filecolor=blue]{hyperref}
\usepackage{color}
\definecolor{dgreen}{rgb}{0.0,0.5,0.0}
\begin{document}

\title{\textcolor{blue}
{Controlled delocalization of electronic states in a multi-strand quasiperiodic  
lattice}}  

\author{Amrita Mukherjee}
\email{amritamukherjee9@yahoo.com}
\author{Atanu Nandy}
\email{atanunandy1989@gmail.com}
\author{Arunava Chakrabarti}
\email{arunava_chakrabarti@yahoo.co.in}
\affiliation{Department of Physics, University of Kalyani, Kalyani,
West Bengal-741 235, India}

\begin{abstract}
Finite strips, composed of a 
periodic stacking of infinite quasiperiodic Fibonacci chains, have been investigated 
in terms of their electronic properties. The system is described by a tight 
binding Hamiltonian. The eigenvalue spectrum of such a multi-strand quasiperiodic network 
is found to be sensitive on the mutual values of the 
intra-strand and inter-strand tunnel hoppings, whose distribution displays a 
unique three-subband self-similar pattern in a parameter subspace. In addition, 
it is observed that special numerical correlations 
between the nearest and the next-nearest neighbor hopping integrals can render a 
substantial part of the energy spectrum {\it absolutely continuous}. Extended, 
Bloch like functions populate the above continuous zones, 
signalling a complete {\it delocalization} 
of single particle states even in such a non-translationally invariant system, and more 
importantly, a phenomenon that can be engineered by tuning the relative strengths of the 
hopping parameters. A commutation relation between the potential and the hopping 
matrices enables us to work out the precise correlation which helps to engineer the 
extended eigenfunctions and determine the band positions at will.
\end{abstract}
\keywords{Delocalization, extended impurity, quasiperiodicity, renormalization.}
\maketitle
\section{Introduction}
\label{sec1}

Recent years have witnessed several interesting variations of the 
classic case of Anderson localization ~\cite{anderson,kramer,abrahams} 
of electronic eigenfunctions in 
disordered systems. Localization of single particle eigenstates in the 
presence of disorder is otherwise, 
an ubiquitous 
phenomenon, extending its realm well beyond the `electronic' scenario, to the 
arena of plasmonic~\cite{christ,ruting}, phononic~\cite{vasseur} or 
polaronic excitations~\cite{barinov}, as well as in the field of photonics~\cite{yablo,john}.
The last example has gained momentum and aroused interest recently after the 
development of the idea of light localization using path-entangled photons~\cite{gilead} and 
the tailoring of partially coherent light~\cite{svozil}. The variants of the phenomenon, 
beginning with the concept of geometrically correlated disorder in the 
distribution of the potentials, the so called {\it random dimer model} 
(RDM)~\cite{dunlap}, or in the overlap integrals in a tight binding description 
~\cite{zhang1}, 
and moving over to the engineering of continuous bands of extended Bloch-like functions 
~\cite{sil,rudo6,arunava1,arunava2,arunava3} in quasi one dimensional disordered or 
quasiperiodic systems, 
thus offer exciting physics and the prospects 
of designing novel devices.

\begin{figure}[ht]
\centering
\includegraphics[clip,width=8.5cm,angle=0]{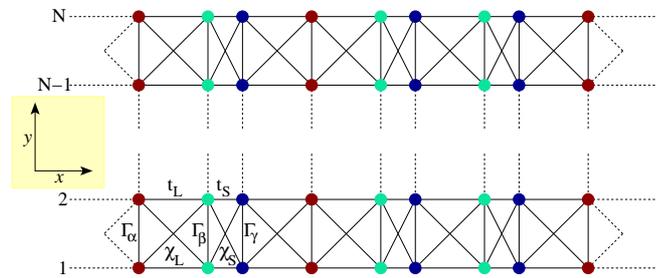}
\caption{(Color online) 
An $N$-strand Fibonacci ladder made up by stacking identical Fibonacci lattices, 
infinitely long in the $x$-direction. The three kinds of vertices, viz., $\alpha$, $\beta$ and 
$\gamma$ are depicted by three different colors red, cyan and blue respectively.
 The second neighbor hoppings
$\chi_L$ and $\chi_S$ are marked distinctly.}
\label{mesh}
\end{figure}

In this communication we explore the possibility of engineering continuous bands of 
extended, Bloch like single particle states in multi-strand ladder networks, 
which on the whole, do not possess any translational invariance. The mesh has  
a finite width in the $y$-direction, but extending to infinity along the $x$-axis. 
Though the problem we address is valid with respect to any kind of 
disordered lattices, we have specifically opted to discuss a quasiperiodic 
network for which an exact analytical attack is possible. 
We work here with a 
mesh formed by infinitely long quasiperiodic Fibonacci chains~\cite{kohmoto},
grown along the $x$-direction and which are  then stacked periodically in the 
transverse $y$-direction. This is illustrated in Fig.~\ref{mesh}. The methodology 
is easily extendable to randomly disordered system of arbitrary width.

It should be appreciated that the 
typical correlated clusters of atomic sites, causing a local resonance
such as in the case of 
the RDM~\cite{dunlap} is absent here, and the spectral 
properties are solely controlled by the interplay of quasiperiodic order along the 
horizontal direction, and the translational invariance (as the system grows in 
the $y$-direction) in the transverse direction. Working out a condition for 
resonance or creation of bands of eigenvalues populated by extended Bloch-like states 
only in such quasi one dimensional systems thus turns out to be non-trivial. 
The `deterministic' growth rule 
of a Fibonacci chain~\cite{kohmoto} allows for an analytical attack on the problem. 
We take advantage of that, obtain the {\it exact} mathematical criteria for generating 
absolutely continuous bands of extended single particle states, and substantiate  
our findings by numerically evaluating the density of states in several cases using 
a real space renormalization group (RSRG) decimation method.

In addition to this, we examine the effect of a {\it line defect} in the form of a linear 
\begin{figure}[ht]
\centering
\includegraphics[clip,width=8.5cm,angle=0]{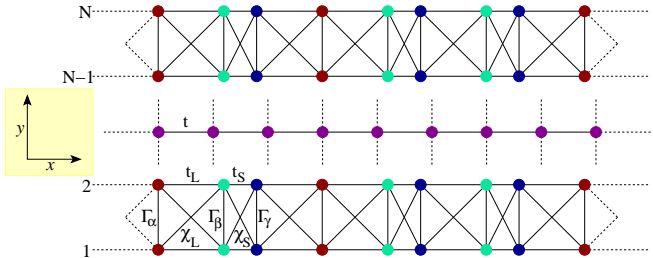}
\caption{(Color online) 
An $N$-strand Fibonacci strip with an embedded 
periodic chain of atoms (magenta colored). All the
atomic sites in the defect chain have the same on-site potential
$\epsilon$ and hopping $t$.}  
\label{defectmesh}
\end{figure}
periodic chain in the bulk of a Fibonacci stack, such as shown in Fig.~\ref{defectmesh}.
The system can be treated in the same mathematical footing as the previous one, and 
gives us a flavor of what influence can a single line defect have on the overall 
energy spectrum of such a quasi-periodic stack. 

The backbone of our analytical attempt is an exact mapping of a 
coupled, quasi-one dimensional multi-strand ladder network into a set of 
totally decoupled linear chains describing the quantum mechanics of a class of 
{\it pseudoparticles}. Such an exact mapping has previously been described in the 
literature in the context of de-localization of single particle states in a ladder-like
geometry~\cite{sil, rudo6} modelling a DNA-like double chain~\cite{sil} or 
a quasi-two dimensional mesh with correlated disorder~\cite{rudo6}.

We find interesting results. For an $N$-strand Fibonacci mesh, it is possible to 
generate absolutely continuous subbands of eigenfunctions by introducing 
appropriate correlation between the numerical values of the parameters of the 
Hamiltonian. For an $N$ strand ladder with $N$ being an odd number, there can be 
just one continuous subband populated by extended states only, at a time. There 
can be $N-1$ different choices for this of course, each choice requiring a different 
correlation 
between the values of the on-site potentials and the nearest and next nearest 
neighbor hopping integrals. For even values of $N$, there can be $N$ such different 
conditions. The spectrum, in all the cases, 
loses the natural three-subband structure of a linear Fibonacci lattice~\cite{kohmoto}.
The inter-strand coupling plays an important role. We have examined the correlation 
between the intra-strand ($t_S$) and inter-strand ($\Gamma$) hopping integrals. The 
distribution of the {\it allowed} combinations of these two displays an interesting 
three subband, self-similar pattern in the $(t_S, \Gamma)$ subspace.

In the latter part of the work we show that the 
incorporation of a linear periodic chain of atoms in the bulk of such a 
quasiperiodic mesh has remarkable influence on the gross spectral behavior. For weak to 
moderate coupling this minimal heterogeneity introduced in the form 
of such single line defect generates regions of absolutely continuous 
bands. The wave functions populating such regions are found to be of extended 
character, as verified by the RSRG recursion relations.

In what follows, we describe the working methodology and the results.
In section~\ref{sec2}, we discuss the Hamiltonian and the basic scheme to engineer 
the continuous subbands in a quasiperiodic mesh. In section~\ref{sec3}, the density of 
states of finite (in $y$-direction) mesh structures are presented, which exhibit the 
continuous bands or subbands. Section~\ref{lineimpurity} deals with the modulation 
of the spectrum as the defect chain is introduced, and in section~\ref{sec5} we draw conclusions.

\section{The Model and the method}
\label{sec2}
\subsection{The Hamiltonian and the Fibonacci mesh}

A prototype geometry representing the kind of system we are interested in, is 
shown in Fig.~\ref{mesh}.  
A single Fibonacci chain is composed of two kinds of `bonds', viz., `long' ($L$) and 
`short' ($S$), and grows following the rule~\cite{kohmoto}, $L \rightarrow LS$, and 
$S \rightarrow L$. A quasi-periodic Fibonacci string in an $i$-th generation 
grows as, 
$G_1 = L$, $G_2 = LS$, $G_3 = LSL$, $G_4 = LSLLS$ and so on. 
In principle, an indefinite number of such arrays can be 
transversely coupled in the $y$-direction periodically or even without periodicity. We 
address the case of a $y$-periodic mesh, consisting of $N$ such linear Fibonacci chains.

The array is modeled by the standard tight binding Hamiltonian 
written in the Wannier basis as,
\begin{widetext}
\begin{equation}
H  = \sum_{i,k} \epsilon_{i}^{k} {{c_{i}}^{k}}^{\dagger} {c_{i}}^{k}
+\sum_{\langle ij \rangle,k} t_{ij}^{k} {c_{i}}^{{k}^{\dagger}} 
{c_{j}}^{k} 
+ \sum_{i,\langle kl \rangle} \Gamma_i {c_{i}}^{{k}^{\dagger}} {c_{i}}^{l} + 
\sum_{\langle ij \rangle, k} {\chi_{ij}}^{k,k+1} {c_{i}}^{{k}^{\dagger}} 
{c_{j}}^{k+1} + 
\sum_{\langle ij \rangle, k} {\chi_{ij}}^{k,k-1} {c_{i}}^{{k}^{\dagger}} {c_{j}}^{k-1}
\end{equation}
\label{hamilton}
\end{widetext}
  
The pairs of indices $(i,j)$ are associated with 
nearest neighbor atomic sites on any particular strand, while 
$k$ and $l$ index represent different strands in the mesh. 
There are three kinds of atomic vertices in each strand, viz.,   
$\alpha$ (red circle), $\beta$ (cyan circle) and 
$\gamma$ (blue circle), depending on whether they are 
flanked by pairs of bonds $L-L$, $L-S$ or $S-L$ respectively. 
The on-site potentials associated with these are $\epsilon_\alpha$, $\epsilon_\beta$ and 
$\epsilon_\gamma$ respectively.
The nearest neighbor hopping integrals in any strand  
are assigned values $t_L$, $t_S$, across the $L$ and the $S$ bonds respectively. 
We incorporate second neighbor hopping between pairs of strands 
across the diagonals in the 
bigger and the smaller rectangular plaquettes as shown, and 
denote them by ${\chi_{ij}}^{k,k\pm 1} = \chi_L$ or  
$\chi_S$ 
respectively according to the geometry. 
The inter-strand tunnel hopping, connecting the 
$i$-th site in the $k$-th strand with the $i$-th site in the $l$-th 
strand ($l = k \pm 1$) is  
$\Gamma_{i}=\Gamma_\alpha$, $\Gamma_\beta$ or $\Gamma_\gamma$ depending on whether 
it connects $\alpha$, $\beta$ or $\gamma$ sites of the neighboring 
strands along the $y$-direction. 
The provision of variation in the values of $\Gamma_i$ ($i=\alpha$, $\beta$ or $\gamma$) 
implies that one can, in principle, discuss the case of quasiperiodically {\it distorted} 
ladder networks as well, bringing in a flavor of geometrical 
disorder and its effect on the energy spectrum of such systems within the same formalism.

The Schr\"{o}dinger equation for the multi-strand network is easily written, in an equivalent 
form using a set of $3N$ difference equations for an $N$ strand ladder. There are 
$N$-equations corresponding to each vertical rung with $\alpha$, $\beta$ or $\gamma$ 
sites residing on it. To avoid complicated equations in the most general form, we 
explicitly write down such difference equations for a three strand ladder network. 
This is enough to bring out the central spirit of the calculations, and a  
generalization to the case of arbitrary $N$ is trivial.  

For a three-strand network, the difference equations for an $\alpha$-rung read, 
\begin{widetext}
\begin{eqnarray}
(E - \epsilon_\alpha) \psi_{i,3} & = & t_L \psi_{i+1,3} + t_L \psi_{i-1,3} 
+ \chi_L \psi_{i+1,2} + \chi_L \psi_{i-1,2} + \Gamma_\alpha \psi_{i,2} \nonumber \\
(E - \epsilon_\alpha) \psi_{i,2} & = & t_L \psi_{i+1,2} + t_L \psi_{i-1,2} 
+ \chi_L \psi_{i+1,3} + \chi_L \psi_{i-1,3} + 
\chi_L \psi_{i+1,1} + \chi_L \psi_{i-1,1} + 
\Gamma_\alpha \psi_{i,3} + \Gamma_\alpha \psi_{i,1} \nonumber \\
(E - \epsilon_\alpha) \psi_{i,1} & = & t_L \psi_{i+1,1} + t_L \psi_{i-1,1} 
+ \chi_L \psi_{i+1,2} + \chi_L \psi_{i-1,2} + \Gamma_\alpha \psi_{i,2} \nonumber \\
\label{diff3str}
\end{eqnarray}
\end{widetext}

Equations for the rungs with $\beta$ - and 
the $\gamma$ sites are,  
\begin{widetext}
\begin{eqnarray}
(E - \epsilon_\beta) \psi_{i,3} & = & t_S \psi_{i+1,3} + t_L \psi_{i-1,3}
+ \chi_S \psi_{i+1,2} + \chi_L \psi_{i-1,2} + \Gamma_\beta \psi_{i,2} \nonumber \\
(E - \epsilon_\beta) \psi_{i,2} & = & t_S \psi_{i+1,2} + t_L \psi_{i-1,2}
+ \chi_S \psi_{i+1,3} + \chi_L \psi_{i-1,3} +
\chi_S \psi_{i+1,1} + \chi_L \psi_{i-1,1} +
\Gamma_\beta \psi_{i,3} + \Gamma_\beta \psi_{i,1} \nonumber \\
(E - \epsilon_\beta) \psi_{i,1} & = & t_S \psi_{i+1,1} + t_L \psi_{i-1,1}
+ \chi_S \psi_{i+1,2} + \chi_L \psi_{i-1,2} + \Gamma_\beta \psi_{i,2} \nonumber \\
\label{diff4str}
\end{eqnarray}
\end{widetext}
and, 
\begin{widetext}
\begin{eqnarray}
(E - \epsilon_\gamma) \psi_{i,3} & = & t_L \psi_{i+1,3} + t_S \psi_{i-1,3}
+ \chi_L \psi_{i+1,2} + \chi_S \psi_{i-1,2} + \Gamma_\gamma \psi_{i,2} \nonumber \\
(E - \epsilon_\gamma) \psi_{i,2} & = & t_L \psi_{i+1,2} + t_S \psi_{i-1,2}
+ \chi_L \psi_{i+1,3} + \chi_S \psi_{i-1,3} +
\chi_L \psi_{i+1,1} + \chi_S \psi_{i-1,1} +
\Gamma_\gamma \psi_{i,3} + \Gamma_\gamma \psi_{i,1} \nonumber \\
(E - \epsilon_\gamma) \psi_{i,1} & = & t_L \psi_{i+1,1} + t_S \psi_{i-1,1}
+ \chi_L \psi_{i+1,2} + \chi_S \psi_{i-1,2} + \Gamma_\gamma \psi_{i,2} \nonumber \\
\label{diff5str}
\end{eqnarray}
\end{widetext}
respectively.

\subsection{Decoupling of the difference equations}
\label{decoupling}

It is simple to recast each of the Eq.~\eqref{diff3str} to Eq.~\eqref{diff5str} 
in a matrix form, viz., 
\begin{widetext}
\begin{equation}
\left [
\left( \begin{array}{cccc}
E & 0 & 0\\
0 & E & 0\\
0 & 0 & E
\end{array}
\right)
-
\left( \begin{array}{cccc}
\epsilon_{i,3} & \Gamma_{32} & 0 \\
\Gamma_{23} & \epsilon_{i,2} & \Gamma_{21} \\
0 & \Gamma_{12} & \epsilon_{i,1} 
\end{array}
\right)
\right ]
\left ( \begin{array}{c}
\psi_{i,3} \\
\psi_{i,2} \\
\psi_{i,1} 
\end{array} \right )
=
\left( \begin{array}{cccc}
t_{i,i+1}^{3} & \chi_{i,i+1}^{32} & 0 \\
\chi_{i,i+1}^{23} & t_{i,i+1}^{2} & \chi_{i,i+1}^{21} \\
0 & \chi_{i,i+1}^{12} & t_{i,i+1}^{1} 
\end{array}
\right )
\left ( \begin{array}{c}
\psi_{i+1,3} \\
\psi_{i+1,2} \\
\psi_{i+1,1}
\end{array} \right )
 +
\left( \begin{array}{cccc}
t_{i,i-1}^{3} & \chi_{i,i-1}^{32} & 0 \\
\chi_{i,i-1}^{23} & t_{i,i-1}^{2} & \chi_{i,i-1}^{21} \\
0 & \chi_{i,i-1}^{12} & t_{i,i-1}^{1}
\end{array}
\right )
\left ( \begin{array}{c}
\psi_{i-1,3} \\
\psi_{i-1,2} \\
\psi_{i-1,1}
\end{array} \right )
\label{matrixdifference}
\end{equation}
\end{widetext}

Here, $\epsilon_{i}$ in the three arms, and along a particular rung are 
$\epsilon_\alpha$, 
$\epsilon_\beta$ or $\epsilon_\gamma$ depending on its status. $t_{i,i \pm 1}^{k}$ 
with $k=1$, $2$ and $3$ will be $t_L$ or $t_S$ depending on the bond in 
the $k$-th arm, 
$\Gamma_{kl}=\Gamma_{lk}$ are $\Gamma_\alpha$, $\Gamma_\beta$ or $\Gamma_\gamma$ 
depending on the sites $\alpha$, $\beta$ or $\gamma$, that are tunnel coupled 
along a vertical rung, 
and $\chi_{i,i \pm 1}^{kl}$ 
will be $\chi_L$ or $\chi_S$ depending on the diagonal that couples the 
$i$-th site of the $k$-th arm ($k=1$, $2$ and $3$) with 
the $i \pm 1$-th site of the $l$-th one $l$ being the arm(s) in the immediate
neighborhood of $k$ in the $y$-direction. 

It is interesting to observe that, with the on-site potentials and the 
nearest and next-nearest neighbor (diagonal) hopping integrals, viz., $t_L$, $t_S$, 
$\chi_L$ and $\chi_S$, the commutator of the potential matrix and the hopping matrix 
$[\tilde{\epsilon_i}, \tilde{t_{ij}}] = 0$, where, 
\begin{displaymath} 
\tilde{\epsilon_i} = 
\left( \begin{array}{cccc}
\epsilon_{i,3} & \Gamma_{32} & 0 \\
\Gamma_{23} & \epsilon_{i,2} & \Gamma_{21} \\
0 & \Gamma_{12} & \epsilon_{i,1} 
\end{array}
\right)
\end{displaymath}
and, 
\begin{displaymath}
\tilde{t_{ij}} = 
\left( \begin{array}{cccc}
t_{i,i+1}^{3} & \chi_{i,i+1}^{32} & 0 \\
\chi_{i,i+1}^{23} & t_{i,i+1}^{2} & \chi_{i,i+1}^{21} \\
0 & \chi_{i,i+1}^{12} & t_{i,i+1}^{1} 
\end{array}
\right )
\end{displaymath}

Taking advantage of this commutation, we can make a change of basis~\cite{sil}, 
by using the relation 
\begin{equation}
\left ( \begin{array}{c}
\phi_{3} \\
\phi_{2} \\
\phi_{1}
\end{array} \right )
= M^{-1} 
\left ( \begin{array}{c}
\psi_{3} \\
\psi_{2} \\
\psi_{1}
\end{array} \right )
\label{basischange}
\end{equation}
The matrix $M$ simultaneously diagonalizes both the potential and the hopping 
matrices $\tilde\epsilon_i$ and 
$\tilde{t_{ij}}$ through a similarity transformation. 
In the new basis $\Phi \equiv (\phi_3$, $\phi_2$, $\phi_1)$ the difference equations
Eq.~\eqref{diff3str} to Eq.~\eqref{diff5str} get decoupled and yield three independent 
sets of difference equations, each of which represents Fibonacci chains, describing 
a kind of {\it pseudoparticles} with states that are 
linear combinations of the old Wannier orbitals $\psi_i$, viz,
\begin{eqnarray}
\phi_3 & = & (-1/2) \psi_3 + (1/2) \psi_1 \nonumber \\
\phi_2 & = & (1/4) \psi_3 - (1/2\sqrt{2}) \psi_2 + (1/4) \psi_1 \nonumber \\
\phi_1 & = & (1/4) \psi_3 + (1/2\sqrt{2}) \psi_2 + (1/4) \psi_1
\label{newbasis}
\end{eqnarray} 
The decoupled, independent, linear equations are:
\begin{align}
& \left[ E - (\epsilon_\alpha - \sqrt{2} \Gamma_\alpha) \right] \phi_{i,3} =
(t_L - \sqrt{2} \chi_L) \phi_{i+1,3} + \nonumber \\
& \qquad \qquad \qquad \qquad \qquad \qquad (t_L - \sqrt{2} \chi_L) \phi_{i-1,3} \nonumber \\
& \left[ E - (\epsilon_\beta - \sqrt{2} \Gamma_\beta) \right] \phi_{i,3} =
(t_S - \sqrt{2} \chi_S) \phi_{i+1,3} + \nonumber \\
& \qquad \qquad \qquad \qquad \qquad \qquad (t_L - \sqrt{2} \chi_L) \phi_{i-1,3} \nonumber \\
& \left[ E - (\epsilon_\gamma - \sqrt{2} \Gamma_\gamma) \right] \phi_{i,3} =
(t_L - \sqrt{2} \chi_L) \phi_{i+1,3} + \nonumber \\
& \qquad \qquad \qquad \qquad \qquad \qquad (t_S - \sqrt{2} \chi_S) \phi_{i-1,3}
\label{decouple1}
\end{align}
\begin{eqnarray}
\left( E - \epsilon_\alpha \right) \phi_{i,2} & = &
t_L \phi_{i+1,2} + t_L \phi_{i-1,2} \nonumber \\
\left( E - \epsilon_\beta \right) \phi_{i,2} & = & 
t_S \phi_{i+1,2} + t_L \phi_{i-1,2} \nonumber \\
\left( E - \epsilon_\gamma \right) \phi_{i,2} & = &
t_L \phi_{i+1,2} + t_S \phi_{i-1,2} 
\label{decouple2}
\end{eqnarray}
\begin{align}
& \left[ E - (\epsilon_\alpha + \sqrt{2} \Gamma_\alpha) \right] \phi_{i,1} =
(t_L + \sqrt{2} \chi_L) \phi_{i+1,1} + \nonumber \\
& \qquad \qquad \qquad \qquad \qquad \qquad (t_L + \sqrt{2} \chi_L) \phi_{i-1,1} \nonumber \\
&\left[ E - (\epsilon_\beta + \sqrt{2} \Gamma_\beta) \right] \phi_{i,1} =  
(t_S + \sqrt{2} \chi_S) \phi_{i+1,1} + \nonumber \\
& \qquad \qquad \qquad \qquad \qquad \qquad (t_L + \sqrt{2} \chi_L) \phi_{i-1,1} \nonumber \\
&\left[ E - (\epsilon_\gamma + \sqrt{2} \Gamma_\gamma) \right] \phi_{i,1} = 
(t_L + \sqrt{2} \chi_L) \phi_{i+1,1} + \nonumber \\ 
& \qquad \qquad \qquad \qquad \qquad \qquad (t_S + \sqrt{2} \chi_S) \phi_{i-1,1} 
\label{decouple3}
\end{align}

A look at the equations Eq.~\eqref{decouple1} $-$ Eq.~\eqref{decouple3} 
\begin{figure}[ht]
\centering
\includegraphics[clip,width=8.5cm,angle=0]{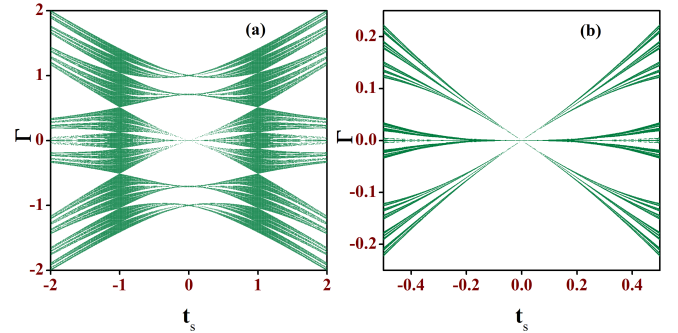}
\caption{(Color online) 
(a) The self similar distribution of the inter-strand hopping against varying 
intra-strand hopping integral $t_S$ for a three strand ladder network
and (b) magnified version of the original spectrum around the central zone. 
We have set 
$\epsilon_i=0$, $t_L=1$ and, $\chi_L=\chi_S=0$. Also,
all the vertical couplings $\Gamma_{\alpha}$, $\Gamma_{\beta}$ and
$\Gamma_{\gamma}$ are set equal to $\Gamma$ for numerical calculation.}
\label{phase}
\end{figure}
reveals that, we now have three independent Fibonacci chains with effective on site 
potentials [ $(\epsilon_\alpha \pm \sqrt{2} \Gamma_\alpha)$, 
$(\epsilon_\beta \pm \sqrt{2} \Gamma_\beta)$, 
$(\epsilon_\gamma \pm \sqrt{2} \Gamma_\gamma)$ ], and 
[$\epsilon_\alpha, \epsilon_\beta, \epsilon_\gamma $].  
Corresponding effective hopping integrals are  $t_L \pm \sqrt{2} \chi_L$, 
$t_S \pm \sqrt{2} \chi_S$, and $t_L$ and $t_S$. Needless to say that, when considered 
individually, each set gives rise to the usual fragmented, Cantor set energy spectrum, 
typical of a one dimensional Fibonacci chain. The actual spectrum of the three strand
ladder is then obtained by convolving the individual spectra. 

Before we end this section we draw the attention of the reader to one pertinent issue 
regarding the nature of the eigenstates of the full Fibonacci mesh that one can 
guess from the decoupled equations.
We should note that, as the wave functions $\phi_i$'s in the new basis are 
each a linear combination of the earlier amplitudes $\psi_i$, localized character of 
any of the $\phi_i$'s will prevail only when every individual $\psi_i$ contributing to 
that particular $\phi_i$ will be localized. On the other hand, if, by certain correlation, 
one can make at least one of the $\psi_i$'s {\it extended} in character, it will render 
the entire linear combination, $\phi_i = \sum_{j}\psi_j$ extended. that is, at least one 
of the independent `channels' will contribute totally transparent states.  
\section{Density of states using the RSRG method}
\label{sec3}
For an arbitrary $N$-strand Fibonacci strip we use the standard RSRG decimation 
using the potential matrices $\tilde{\epsilon_i}$ and the hopping matrices 
$\tilde{t_{ij}}$. The decimation is implemented by `folding' the multi-strand strip 
using the growth rule in the reverse direction, viz., $LS \rightarrow L'$ and 
$L \rightarrow S'$. The RSRG recursion relations are easily obtained as, 
\begin{widetext}
\begin{eqnarray}
\tilde{\epsilon}_{\alpha,n+1} & = & \tilde{\epsilon}_{\gamma,n} + 
\tilde{t}_{S,n}^{T}(E.I -\tilde{\epsilon}_{\beta,n})^{-1} \tilde{t}_{S,n} +
\tilde{t}_{L,n} (E.I - \tilde{\epsilon}_{\beta,n})^{-1} \tilde{t}_{L,n}^{T} \nonumber \\
\tilde{\epsilon}_{\beta,n+1} & = & \tilde{\epsilon}_{\gamma,n} + 
\tilde{t}_{S,n}^{T} (E.I - \tilde{\epsilon}_{\beta,n})^{-1} \tilde{t}_{S,n} \nonumber \\
\tilde{\epsilon}_{\gamma,n+1} & = & \tilde{\epsilon}_{\alpha,n} +      
\tilde{t}_{L,n} (E.I - \tilde{\epsilon}_{\beta,n})^{-1} \tilde{t}_{L,n}^{T} \nonumber \\
\tilde{t}_{L,n+1} & = & \tilde{t}_{L,n} (E.I - \tilde{\epsilon}_{\beta,n})^{-1} 
\tilde{t}_{S,n} \nonumber \\
\tilde{t}_{S,n+1} & = & \tilde{t}_{L,n}
\label{recursion}
\end{eqnarray}
\end{widetext}
With a small imaginary part added to the energy $E$, the matrix elements of the 
hopping matrix $\tilde{t_L}$ and $\tilde{t_S}$ flow to zero after a finite number 
of iterations. The Green's function matrix $\tilde{G}(E)$ is then obtained from the 
equation $\tilde{G}(E+i\eta) = [(E + i \eta).I - \epsilon_i^{*}]^{-1}$, where the
subscript $i=\alpha$, 
$\beta$ or $\gamma$, and the `asterix' denotes the {\it fixed point} value of the 
corresponding on-site potential. The diagonal elements of the Green's function matrix 
provide the density of states at the vertices on an $\alpha$, $\beta$ or a 
$\gamma$ strand.  
\subsubsection{The role of interstrand coupling}
We make an interesting observation to begin with. We fix $t_L=1$, and 
look for the nature of distribution of the mutual values of the intra-strand 
hopping $t_S$ and the inter-strand one, that is $\Gamma_i$, with $\chi_L=\chi_S=0$ for the 
time being. For simplicity we set $\Gamma_\alpha=\Gamma_\beta=\Gamma_\gamma=\Gamma$, and 
vary both $t_S$ and $\Gamma_i$ within $[-2,2]$ in units of $t_L$. 
The changing values of $\Gamma$, in a naive way, correspond to the proximity of 
the strands of the ladder.
\begin{figure}[ht]
\centering
\includegraphics[clip,width=7cm,angle=0]{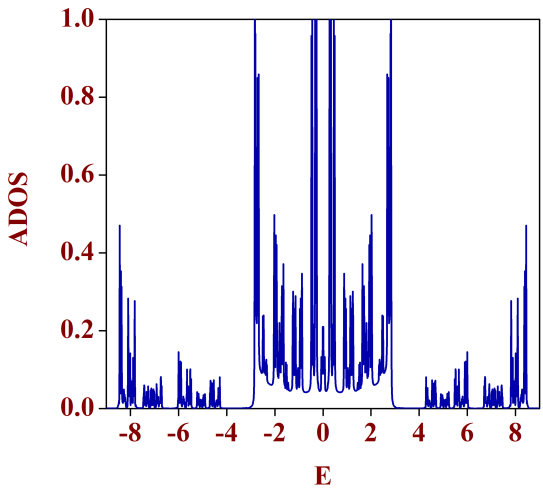}
\caption{(Color online) 
Average density of states of a $3$-strand Fibonacci strip with correlated parameters.
$\chi_L = 1 + 1/\sqrt{2}$ and $\chi_S = 1 + \sqrt{2}$.
 We have chosen $\epsilon_i=0$ for 
$i = \alpha$, $\beta$, $\gamma$ and $\delta$, and $t_L=1$ and $t_S=2$. Energy 
is measured in units of $t_L$.}
\label{dos3strand}
\end{figure}
The plot in 
Fig.~\ref{phase} depicts the $(\Gamma$, $t_S)$ parameter subspace for which we 
get a non-zero density of states of the full three-strand Fibonacci ladder. The density 
of states has been obtained by using the recursion relations Eq.~\ref{recursion}.
It is interesting to observe that the distribution of the inter-strand tunnel hopping 
displays a {\it three subband}, self similar, fractal character.   
In Fig.~\ref{phase}(b) we have blown up an area around the 
origin to highlight the trifurcating character of the distribution.

\subsubsection{Engineering absolutely continuous bands and extended wavefunctions}
The strength of the decoupling scheme lies in its ability to engineer 
absolutely continuous energy bands even in such a system which doesn't have  any translational 
invariance. We justify the claim by pointing out to the fact that, for example, 
in Eq.~\eqref{decouple1}, if one sets $\epsilon_\alpha - \sqrt{2} \Gamma_\alpha = 
\epsilon_\beta - \sqrt{2} \Gamma_\beta = \epsilon_\gamma - \sqrt{2} \Gamma_\gamma$, and 
$t_L - \sqrt{2} \chi_L = t_S - \sqrt{2} \chi_S$, then the set of Eq,~\eqref{decouple1} 
represents a perfectly ordered lattice. The energy eigenvalues 
of this {\it effectively periodic} chain form an absolutely continuous 
band with the pseudoparticle states {\it extended}, Bloch-like. The band is centered at 
$E = \epsilon_\alpha - \sqrt{2} \Gamma_L$, and extends from 
$E = \epsilon_\alpha - \sqrt{2} \Gamma_L - 2 (t_L - \sqrt{2} \chi_L)$ to 
$E = \epsilon_\alpha - \sqrt{2} \Gamma_L + 2 (t_L - \sqrt{2} \chi_L)$. 
It is to be noted that, such a correlation does not restrict the individual 
values of the potentials $\epsilon_i$, and 
the inter-strand hopping integrals $\Gamma_i$ from being chosen from any desired 
distribution. We are only demanding that the difference $\epsilon_i-\sqrt{2}\Gamma_i$ 
be kept constant. Similar is the case with the difference 
$t^k_{ij} - \sqrt{2}\chi^{k,k\pm 1}_{ij}$.
Similar observation is made with respect to Eq.~\eqref{decouple3}, where the correlations 
needed for the creation of an absolutely continuous band are, 
$\epsilon_\alpha  +\sqrt{2} \Gamma_\alpha = \epsilon_\beta + \sqrt{2} \Gamma_\beta = $
$\epsilon_\gamma+\sqrt{2} \Gamma_\gamma$, and 
$t_L + \sqrt{2} \chi_L = t_S + \sqrt{2} \chi_S$.

In Fig.~\ref{dos3strand} we show the density of states of a three-strand 
Fibonacci strip with $\epsilon_i=0$, and 
with the above correlations respectively. We have taken,
without any loss of generality, $\Gamma_\alpha=\Gamma_\beta=
\Gamma_\gamma=0$, just to set the center of the spectrum at $E=0$, and the inter-strand 
coupling survives through the diagonal hopping integrals alone. 
It is easily understood that 
at a time only one of the two equations, 
viz., Eq.~\eqref{decouple1} or Eq.~\eqref{decouple3} 
can be made to contribute an absolutely continuous band populated by extended 
Bloch like states only. The remaining two equations will populate the spectrum with 
critical eigenstates, characteristic of a Fibonacci chain. 
However, if some of the critical states 
happen to be occupying part of the spectrum that falls within the band of extended 
wave functions, then they will lose their {\it critical} identity and 
become a part of the extended family. 
\begin{figure}[ht]
\centering
\includegraphics[clip,width=7cm,angle=0]{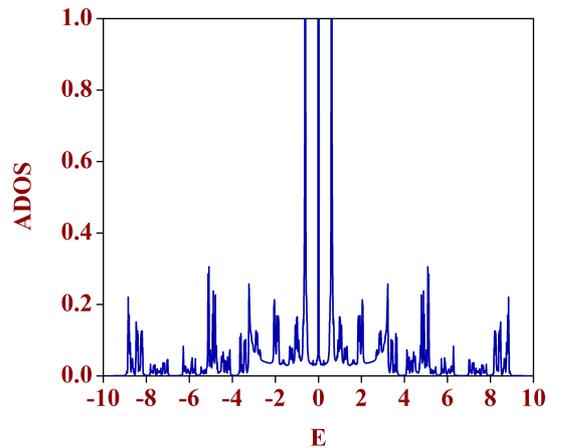}
\caption{(Color online) 
Average density of states of a $4$-strand Fibonacci strip with correlated parameters.
$\chi_L = 1 + 2/(\sqrt{5}+1)$  and $\chi_S = 1 + 4/(\sqrt{5}+1)$.
We have chosen $\epsilon_i=0$ for 
$i = \alpha$, $\beta$, $\gamma$ and $\delta$, and $t_L=1$ and $t_S=2$. Energy 
is measured in units of $t_L$.}
\label{dos4strand}
\end{figure}
This is precisely what happens in 
Fig.~\ref{dos3strand}. We have set the correlation between the nearest and next 
nearest neighbor (along the diagonals) hopping integrals as, $t_L - \sqrt{2}\chi_L 
= t_S - \sqrt{2}\chi_S$. With $t_L=1$, $t_S=2$, $\chi_L=1+1/\sqrt{2}$ and 
$\chi_S=1+\sqrt{2}$, Eq.~\eqref{decouple1} represents a periodic chain of 
atomic sites with absolutely continuous band of extended states lying between 
$-2\sqrt{2} \le E \le 2\sqrt{2}$. Needless to say that, the two other equations, viz. 
Eq.~\eqref{decouple2} and Eq.~\eqref{decouple3} still represent two independent 
Fibonacci chains giving rise to `critical' eigenstates. The full spectrum, as obtained 
from the RSRG recursion relations and the Green's functions reproduce the absolute
\begin{figure*}[ht]
\centering
\includegraphics[clip,width=16cm,angle=0]{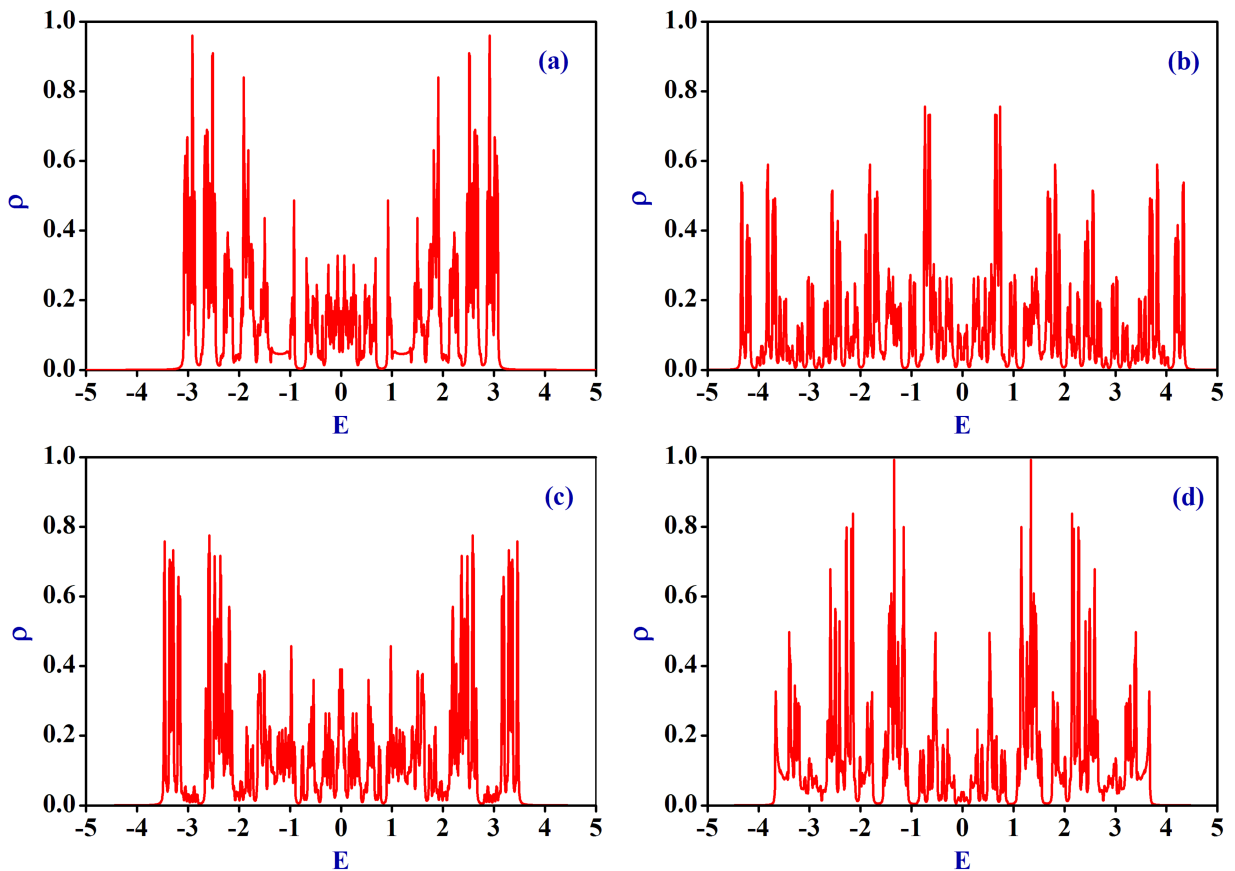}
\caption{(Color online) 
Density of states of a $5$-strand Fibonacci strip with an embedded periodic chain 
in the $3$rd strand. The values of the parameters are $\epsilon_i = 0, i = 
\alpha$, $\beta$ and $\gamma$, $t_L = 1$, $t_S = 2$, $\chi_L = 0 = \chi_S$. 
The vertical couplings are (a) $\Gamma_\alpha = \Gamma_\beta = \Gamma_\gamma = 0.2$, 
(b) $\Gamma_\alpha = \Gamma_\beta = \Gamma_\gamma = 1$,
(c) $\Gamma_\alpha = 0.2$, $\Gamma_\beta = \Gamma_\gamma = 0.5$ and
(d) $\Gamma_\alpha = 1$, $\Gamma_\beta = \Gamma_\gamma = 0.5$. All the atomic
sites of the
extended defect chain have the same on-site potential
$\epsilon = 0$ and same hopping integral $t = 1$.}
\label{defectdos}
\end{figure*}
continuum exactly over the energy regime, as extracted from the 
decoupled Eq.~\eqref{decouple1}.

The `extendedness' of the states has been 
verified by picking up arbitrary energy eigenvalues from inside the  
central portion ($-3 < E < 3$) in Fig.~\ref{dos3strand}, 
and observing the flow of the hopping integrals under successive 
RSRG iteration steps, with the imaginary part of the energy set equal to zero. 
For all such energy eigenvalues the hopping integrals kept on oscillating without 
converging to zero for an arbitrary number of loops. This is an indication that the 
corresponding Wannier orbitals have finite overlap over arbitrarily large distances - 
a distinct signature of the extended character of the eigenfunction.

A similar test has been carried out to examine the nature of eigenstates 
populating the subbands at the flanks. We have encountered a whole bunch of 
eigenvalues for which the hopping integrals ultimately flow to zero,
but only after a large 
number of RSRG iterations. This is suggestive of the fact that the corresponding 
eigenfunctions at least have very large localization lengths,
if not `extended' (if we consider a practical situation).
It should be mentioned that, we have used the 
decoupled equations only to extract the region of the central (in this case) continuous subband. The density 
of states presented in every figure is obtained by using the RSRG method on the 
full multi-strand Fibonacci mesh.

Using the same set of values for the on-site potentials $\epsilon_i$ and the inter-strand 
vertical hopping $\Gamma_i$, $i=\alpha$, $\beta$, and $\gamma$, we present the 
DOS spectrum for the $4$-strand Fibonacci ladder in Fig.~\ref{dos4strand}. Following the 
prescription laid out above, it is now straightforward to understand that the change 
of basis decouples this relatively complicated system into four independent 
linear chains, each representing the difference equation for the  `mixed' states, 
viz., $\phi_i = \sum_N \psi_{i,N}$. Each individual equation represents, as before, 
`pure' Fibonacci quasiperiodic chains with both the effective on-site potentials 
and the hopping integrals arranged in Fibonacci sequence, and the spectrum 
offered by each one of them is a fragmented Cantor set, with the wave functions 
`critical' in general, exhibiting a power-law localization and the usual multifractality.
In Fig.~\ref{dos4strand} the spectrum is a mixed one, the central part being 
composed of an absolutely continuous band populated by extended eigenstates only. 
This continuum is a result of the correlation between $t_L$, $\chi_L$ and 
$t_S$, $\chi_S$ in a manner similar to the case of a three-strand ladder. The 
central part is flanked by fragmented subbands populated by localized eigenfunctions. 

We have tested the extendedness of the wavefunctions in this case also 
in regions which appear 
to be continuous by observing the flow of the hopping integrals under the RSRG steps. 
The conclusions are the same as in the three-strand case discussed earlier.
\section{The role of an `extended impurity'}
\label{lineimpurity}
We now examine the effect of an {\it extended impurity} segment, in the shape of an 
infinitely long periodically ordered chain embedded in the bulk of a multi-strand 
Fibonacci ladder. We refer to Fig.~\ref{defectmesh}. No special correlations between the 
numerical values of the hopping integrals is introduced. 
This is an interesting situation when heterogeneity in the system is introduced 
at the minimal level. The influence of coupling transport channels which exhibit 
completely different localization properties has already provided exciting results 
~\cite{markus}, and the present study draws inspiration out of this.

We present the results of the DOS for a five-strand ladder network where the 
central strand is an ordered chain, extending up to infinity along the $x$-direction.
We have not gone for any special correlation between the hopping integrals as in 
the earlier situations. Yet, for weak (compared to $t_S$) tunnel hopping 
$\Gamma_i$, the panel (a) in Fig.~\ref{defectdos} brings out the presence of a 
broad continua in segments in the DOS spectrum. Such zones turn out to be populated by extended 
eigenfunctions only, as we have tested by observing the flow of the hopping integrals 
under successive steps of renormalization. The flanks of the central continuum 
are populated by localized states, for which the hopping integrals, under RSRG 
iteration, flow to zero after a finite (in some cases a remarkably large number even) 
number of iterations. With increasing values of $\Gamma_i$ the feature still persists, 
unless, for a large enough value, the spectrum fragments into sharply localized 
subclusters with the hopping integrals flowing to zero after a 
nominal number of RSRG iteration. 
However, the typical trifurcating pattern observed in the DOS of a 
one dimensional Fibonacci chain (transfer model) is absent. 
\section{Concluding remarks}
\label{sec5}
We have studied the electronic properties of a quasi-one dimensional 
quasiperiodic lattice in the form of a multi-strand ladder network. We have 
specifically chosen a Fibonacci seqeunce to generate such a structure and have 
investigated the spectral characters by evaluating the densities of states 
using a real space renormalization group formalism. A commutation relation between the 
potential and the hopping matrices has been exploited to work out special 
correlations between the numerical values of various hopping integrals by virtue 
of which one can create absolutely continuous subbands of extended eigenstates only. 
This aspect provides a non-trivial variation over the canonical case of Anderson 
localization. Finally, the insertion of a line impurity is shown to lead to a gross 
change in the spectral character, leading to the generation of extended eigenfunctions 
as well.
\begin{acknowledgments}
A.N. is thankful to UGC, India for providing a research fellowship [Award Letter No.-
F.$17-81/2008$(SA - I)]. 
\end{acknowledgments} 

\end{document}